\documentclass[conference]{IEEEtran}
\usepackage{cite}
\ifCLASSINFOpdf
\usepackage{graphicx}
\graphicspath{{./images/}}
\usepackage{xcolor}
\DeclareGraphicsExtensions{.pdf,.jpeg,.png}
\usepackage{amsmath}
\usepackage{array}
\usepackage[caption=false,font=normalsize,labelfont=sf,textfont=sf]{subfig}
\usepackage{stfloats}
\usepackage{float}
\usepackage{url}
\newcommand{\tildebf}[1]{\Tilde{\mathbf{#1}}}

\begin{document}

\title{Interference and Rate Analysis of Multinumerology NOMA}

\author{\IEEEauthorblockN{Stephen McWade\IEEEauthorrefmark{1},
Mark F. Flanagan\IEEEauthorrefmark{1},
Lei Zhang\IEEEauthorrefmark{2} and 
Arman Farhang\IEEEauthorrefmark{3}}
\IEEEauthorblockA{\IEEEauthorrefmark{1}School of Electrical, Electronic and Communications Engineering,
University College Dublin,
Dublin, Ireland \\}
\IEEEauthorblockA{\IEEEauthorrefmark{2}School of Engineering, University of Glasgow, Glasgow G12 8QQ, U.K.\\}
\IEEEauthorblockA{\IEEEauthorrefmark{3}Department of Electronic Engineering,
National University of Ireland Maynooth,
Maynooth, Ireland\\
Email: stephen.mcwade@ucdconnect.ie, mark.flanagan@ieee.org, Lei.Zhang@glasgow.ac.uk, arman.farhang@mu.ie}
}
\maketitle
\begin{abstract}
5G communication systems and beyond are envisioned to support an extremely diverse set of use cases with different performance requirements. These different requirements necessitate the use of different numerologies for increased flexibility. 
Non-orthogonal multiple access (NOMA) can potentially attain this flexibility by superimposing user signals while offering improved spectral efficiency (SE). However, users with different numerologies have different symbol durations. When combined with NOMA, this changes the nature of the interference the users impose on each other. This paper investigates a multinumerology NOMA (MN-NOMA) scheme using successive interference cancellation (SIC) as an enabler for coexistence of users with with different numerologies. Analytical expressions for the inter-numerology interference (INI) experienced by each user at the receiver are derived, where mean-squared error (MSE) is the metric used to quantify INI. Using the MSE expressions, we analytically derive achievable rates for each user in the MN-NOMA system. These expressions are then evaluated and used to compare the SE performance of MN-NOMA with that of its single-numerology counterpart. The proposed scheme can achieve the desired flexibility in supporting diverse use cases in future wireless networks. The scheme also gains the SE benefits of NOMA compared to both multinumerology and single numerology orthogonal multiple access (OMA) schemes.

\end{abstract}


\IEEEpeerreviewmaketitle

\section{Introduction}

5G mobile networks are envisioned to have the flexibility to support a wide range of different services. These services have been broadly categorized into three main usage scenarios: enhanced mobile broadband (eMBB), ultra-reliable and low-latency communications (URLLC) and massive machine-type communications (mMTC) \cite{2015IMTV}. All of these use cases have different performance requirements. For example, mMTC may require a narrow subcarrier spacing to support delay-tolerant devices. On the other hand, URLLC has much more stringent latency requirements and would require a wide subcarrier spacing (thus a smaller symbol duration) \cite{Zhang2017}. It is clear then that a one-size-fits-all numerology design, as is the case in 4G Long Term Evolution (LTE), may not be able to provide the desired flexibility required to support this diverse set of services. In addition, designing separate radios for separate services is not viable as the operation and management of the systems would be extremely complex \cite{Zhang2017}.

To support these services in 5G and beyond, one solution is to separate the system bandwidth into smaller bandwidth parts (BWPs) with each BWP having specifically designed physical layer parameters to meet the stringent requirements of different services. According to 3GPP, orthogonal frequency division multiplexing (OFDM) is still the base waveform for 5G \cite{3GPP_2016}. OFDM subcarriers with differing numerologies, e.g. different subcarrier spacings and cyclic prefix (CP) lengths, will no longer be orthogonal to each other. This makes an OMA solution, where the co-existing numerologies are adjacent to each other and non-overlapping in the frequency domain, difficult if not impossible. Plenty of research into an OMA approach to this problem has been performed so far. The authors of \cite{demmer_2018} provided a MSE analysis of the interference caused by adjacent numerologies and used it to set a minimum guard band between numerologies to achieve a target MSE. The authors of \cite{Kihero2018} investigated the primary factors that contribute to inter-numerology interference (INI) for users in adjacent BWPs. The authors of \cite{zhang_2018} derived expressions for INI and used them as the basis for a novel interference cancellation scheme for windowed OFDM. 

An alternative to the OMA approach is to use non-orthogonal multiple access (NOMA). The basic concept of NOMA is well-established. Users share time and frequency resources and are multiplexed in another domain such as the power domain or the code domain \cite{lu_noma_2017}, and superposition coding and successive interference cancellation (SIC) are used to accurately decode the user signals. NOMA has many desirable performance attributes when compared to OMA, particularly with regard to its improved spectral efficiency \cite{lu_noma_2017}. However, conventional NOMA assumes that users have the same subcarrier spacing \cite{popovski2018}. Given that future networks will need to support services with heterogeneous subcarrier spacing, there is a clear motivation for using NOMA for users with different numerologies to improve the spectral efficiency of multiuser systems. 

Conventional NOMA techniques are already well-investigated \cite{lu_noma_2017}, but to the best of our knowledge, there is little research on NOMA in a multinumerology context so far. The authors of \cite{popovski2018} outline orthogonal slicing and non-orthogonal slicing of different services from the communication-theoretic point of view. However, this work does not consider the interference caused by the different numerologies of the heterogeneous services. 

The authors of \cite{Abusabah2018NOMAFM} outline a NOMA scheme with different users using different numerologies. However, the system outlined in \cite{Abusabah2018NOMAFM} does not use multiple numerologies to achieve greater system flexibility. Instead, different subcarrier spacings are used to reduce the level of co-channel interference between users and increase rate allocation fairness. The authors of \cite{Abusabah2018NOMAFM} also only consider large scale block fading in their channel model. Moreover, the system used in \cite{Abusabah2018NOMAFM} shows a SE loss compared to conventional NOMA systems.

Against this background, the contributions of this paper are as follows:

\begin{itemize}
\item We propose a  multinumerology uplink NOMA scheme wherein users having different numerologies transmit over multipath fading channels and are decoded via SIC at the receiver.
\item  For the proposed scheme, we provide a derivation and analysis of the INI, in terms of MSE, caused by the different numerologies coexisting under multipath fading channel conditions in a NOMA system.
\item We analyze the achievable rates for each user in a multinumerology NOMA scenario using the INI analysis. 
\item We show that a multinumerology NOMA system provides superior SE to a multinumerology OMA system and that there is no SE loss compared to SN-NOMA.  
\end{itemize}

The rest of this paper is organized as follows. In Section II we describe the system model. In Section III we derive analytical expressions for the INI and corresponding achievable rates for users with different numerologies. In Section IV we provide numerical results and discussion based on the analysis from the previous section. Finally, Section V concludes the paper.

\subsubsection*{Notations} Superscripts ${(\cdot)^{\rm{T}}}$ and ${(\cdot)^{\rm{H}}}$ denote transpose and Hermitian transpose, respectively. Boldface lower-case characters are used to denote vectors and boldface upper-case characters are used to denote matrices. diag$(\mathbf{X})$ is a column vector whose elements include the main diagonal of the matrix $\mathbf{X}$ and $\otimes$ represents Kronecker product. The superscript ${^{(i)}}$ is used to denote the $i$-th user. The $p\times{p}$ identity matrix is denoted by $\mathbf{I}_p$. An all zeros matrix of size $p\times{q}$ is denoted by $\mathbf{0}_{p\times{q}}$.

\section{System Model}
\begin{figure}[t]
    \begin{subfloat}
    \centering
    \includegraphics[width=\columnwidth]{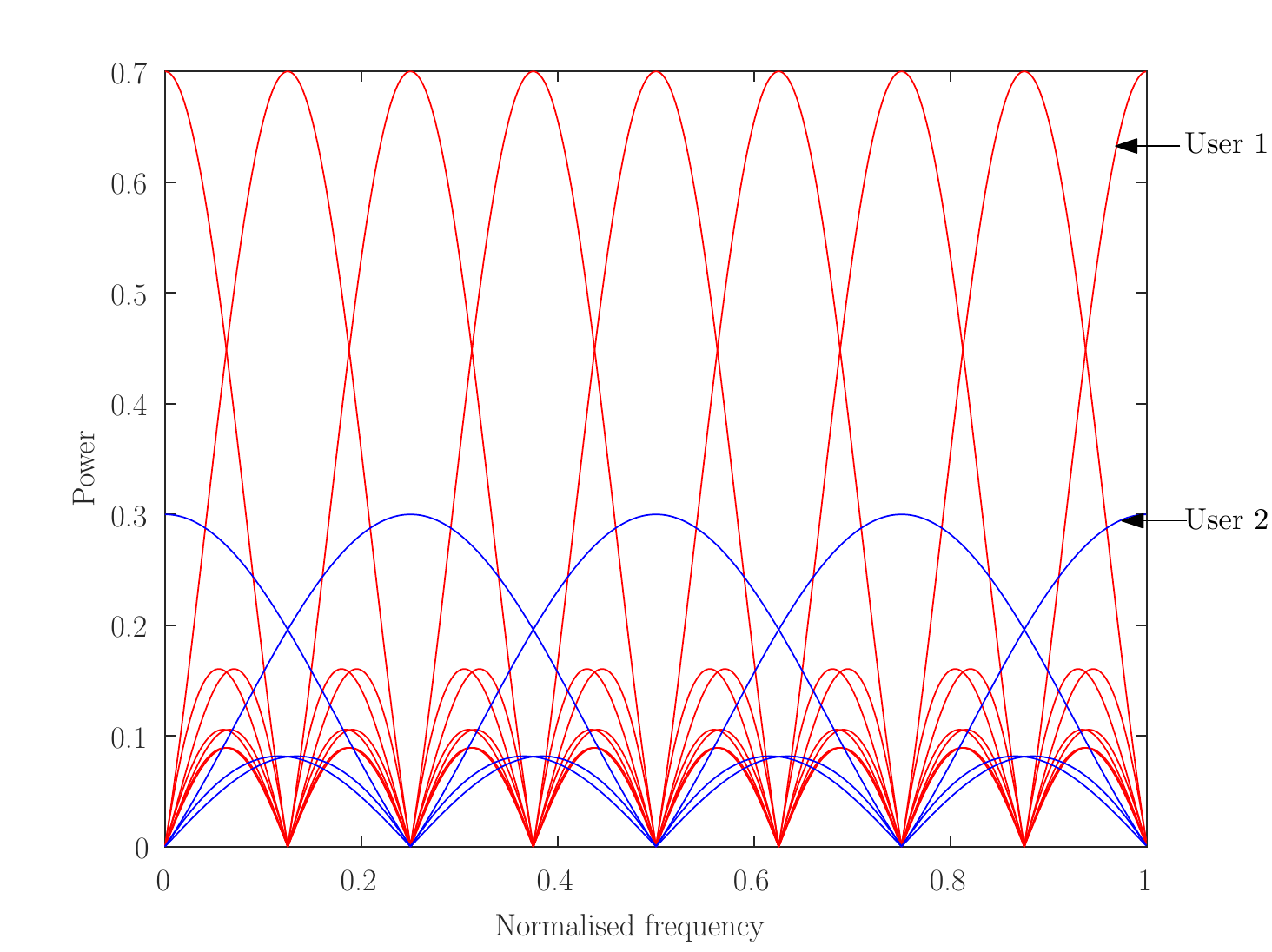}
    \caption{\label{fig:1} Illustration of subcarriers for the proposed MN-NOMA system for the case $q=2$.}
    \end{subfloat}
    \begin{subfloat}
    \centering
    \includegraphics[width=\columnwidth]{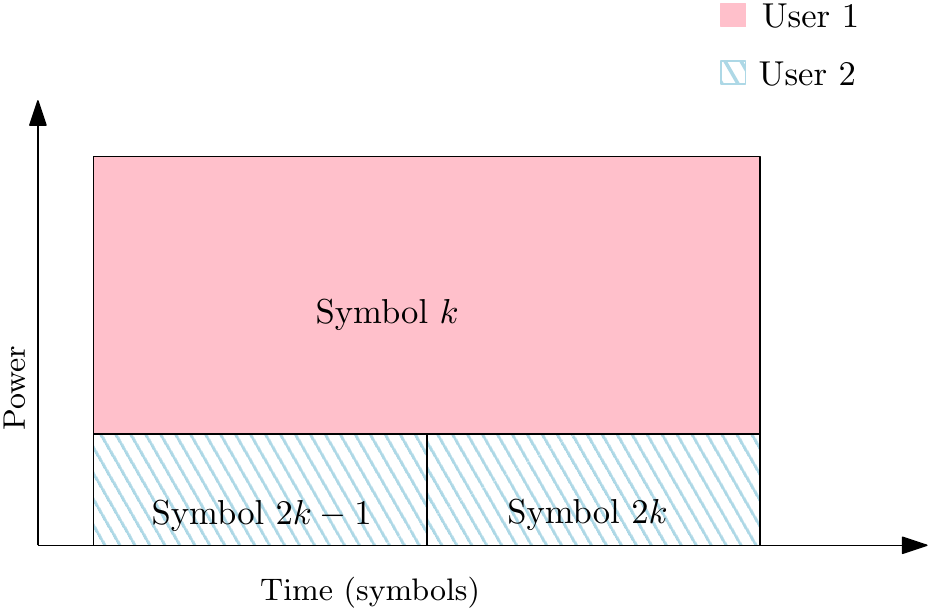}
    \caption{\label{fig:1.2} Illustration of time and power domain multiplexing for the proposed MN-NOMA system for the case $q=2$.}
    \end{subfloat}
\end{figure}

For clarity and simplicity of the derivations, without loss of generality, this paper considers a 2-user uplink setup with each user using a different numerology. Each user $i \in \{ 1,2 \}$ uses a subcarrier spacing $f^{(i)}$; these are related by $\Delta{}f^{(2)} = q\Delta{}f^{(1)}$, where \(q = 2^{\mu}\), \(\mu \in \{0,1,2,3...\}\), as per 3GPP \cite{3gpp_TS38211}. Fig. \ref{fig:1} shows an illustration of the subcarriers of user 1 and user 2 for the case $q = 2$. The symbol duration of the users are therefore related by $T^{(2)} = T^{(1)}/q$. Fig. \ref{fig:1.2} shows an illustration of the time-domain symbols of each user after power domain multiplexing, with $q = 2$. 
User 1 uses a unitary DFT matrix of size $N^{(1)}$ for OFDM modulation. This is $q$ times larger than the size $N^{(2)}$ unitary DFT matrix for user 2. The numerology of user $i$ $ \in \{1, 2\}$ has a corresponding CP length $N_{\rm{CP}}^{(i)}$ which is  scaled for each possible numerology so that there is no SE loss. The total symbol length is $L^{(i)} = N^{(i)} + N_{\rm{CP}}^{(i)}$ . 
\subsection{Transmitted signal for User $i$ $\in \{1, 2\}$}
Each user's transmitted signal is constructed as a conventional CP-OFDM signal. The \(k\)th OFDM symbol corresponds to the \(N_{\rm{act}}^{(i)}\ \times 1\) vector of data-bearing symbols on the active subcarriers,

\begin{equation}
\mathbf{d}_k^{(i)} = \left[d_{k,0}^{(i)}, d_{k,1}^{(i)}, \dots  ,d_{k,N_{\rm{act}}^{(i)}}^{(i)}\right]^{\rm{T}}, 
\label{eq:1}
\end{equation}
where each \(d_{k,n}^{(i)}\) is an individual data-bearing symbol of unit energy. \(N_{\rm{act}}^{(i)}\) is the number of data-bearing subcarriers for user $i$.
The \(k\)th OFDM symbol is constructed as
\begin{equation}
\mathbf{x}_k^{(i)} = {\mathbf{A}_{\rm{cp}}^{(i)}}({\mathbf{F}^{(i)})^{\rm{H}}}{\mathbf{d}_k^{(i)}},\label{eq:2}
\end{equation}
where \(\mathbf{A}_{\rm{cp}}^{(i)} = \left[\mathbf{I}_{\rm{cp}}^{(i)}, \mathbf{I}_{N^{(i)}}\right]^{\rm{T}}\) is a matrix which adds the CP to the symbol and $\mathbf{I}_{{\rm{cp}}}^{(i)}$ is composed of the final $N_{\rm{CP}}^{(i)}$ columns of $\mathbf{I}_{N^{(i)}}$.
\(\mathbf{F}^{(i)}\) is an $N_{\rm{act}}^{(i)}\times N^{(i)}$ sub-matrix of the \(N^{(i)}\)-point unitary DFT matrix, obtained by including the rows which correspond to the active subcarriers.

\subsection{Power Domain Multiplexing of User Signals}
The transmitted signals for user 1 and user 2 are given by 

\begin{equation}
    \mathbf{s}^{(1)}_k = \sqrt{p^{(1)}}\mathbf{x}_k^{(1)},
    \label{eq:3}
\end{equation}
and
\begin{equation}
    \mathbf{s}^{(2)}_k = \sqrt{p^{(2)}}\tildebf{x}_k^{(2)},
    \label{eq:3.1}
\end{equation}
respectively, where \(p^{(i)}\) is the transmit power of user \(i\). The system has a total power budget of $P = p^{(1)}+p^{(2)}$. 

The vector \(\tildebf{x}_k^{(2)}\) is the concatenation of the \(q\) OFDM symbols from user 2 which occur during the duration of symbol \(\mathbf{x}_k^{(1)}\) from user 1, and is given by

\begin{equation}
    \tildebf{x}_k^{(2)} = \left[(\mathbf{x}_{k, 1}^{(2)})^{\rm{T}}, (\mathbf{x}_{k,2}^{(2)})^{\rm{T}}, \dots , (\mathbf{x}_{k, q}^{(2)})^{\rm{T}} \right]^{\rm{T}},
    \label{eq:4}
\end{equation}
where $\mathbf{x}_{k, m}^{(2)}$ represents the $m$-th symbol of user 2 which overlaps with the $k$-th symbol of user 1 where $1 \leq m \leq q$. This concatenated symbol can also be expressed as

\begin{equation}
    \tildebf{x}_k^{(2)} =\left[\mathbf{I}_q\otimes(\mathbf{A}_{\rm{cp}}^{(2)}({\mathbf{F}^{(2)})^{\rm{H}}})\right]\tildebf{d}_k^{(2)},
    \label{eq:5}
\end{equation}
where 
$$\tildebf{d}_k^{(2)} = \left[(\mathbf{d}_{k, 1}^{(2)})^{\rm{T}}, (\mathbf{d}_{k,2}^{(2)})^{\rm{T}}, \dots , (\mathbf{d}_{k, q}^{(2)})^{\rm{T}} \right]^{\rm{T}}$$
is a vector of concatenated data-bearing symbols corresponding to $\tildebf{x}_k^{(2)}$.

\subsection{Channel}
We assume that the time-domain channel impulse response for user $i$ can be written as $\mathbf{h}^{(i)} = \left[h^{(i)}_0, \dots ,h^{(i)}_{N_{ch}}\right]^{\rm{T}}$ where $N_{ch}$ denotes the length of the channel. The received signal can be expressed as

\begin{equation}
    \mathbf{r}_k = \mathbf{H}^{(1)}\mathbf{s}^{(1)}_k + \mathbf{H}^{(2)}\mathbf{s}^{(2)}_k + \mathbf{w}_k.
    \label{eq:6}
\end{equation}
where $\mathbf{H}^{(i)}$ is the Toeplitz channel matrix of user $i$ with first columns equal to $$\left[(\mathbf{h}^{(i)})^{\rm{T}}, \mathbf{0}_{1\times(L^{(i)}-N_{ch}-1)}\right]$$ and first row equal to $$\left[h_0^{(i)}, \mathbf{0}_{(1)\times(L^{(i)}-1)}\right].$$ The vector $\mathbf{w}_k \sim \mathcal{CN}(\mathbf{0}, N_0\mathbf{I})$ represents additive white Gaussian noise (AWGN) of power spectral density $N_0$. The system's signal to noise ratio is defined as $\mathrm{SNR} = {P}/{N_0}$.

\subsection{SIC Receiver}
The receiver decodes the signals using SIC. There exist two possible orderings of the users for the SIC process, one where user 1 is decoded first and one where user 2 is decoded first. The user that is decoded first experiences interference from the other user and the order of SIC affects the interference experienced. The natural asymmetry between the MN-NOMA users means that it is necessary to analyze the interference for both user orderings separately.

\subsubsection*{Ordering 1 - SIC where user 1 is decoded first}

The received signal is processed via
\begin{equation}
    \mathbf{y}_k^{(1)} = \mathbf{F}^{(1)}\mathbf{R}_{\rm{cp}}^{(1)}\mathbf{r}_k,
    \label{eq:7}
\end{equation}
where $\mathbf{R}_{\rm{cp}}^{(1)}$ is a matrix for removing the CP of user 1. Using (\ref{eq:3}), (\ref{eq:3.1}) and (\ref{eq:6}) then yields
\begin{equation}
    \begin{split}
    \mathbf{y}_k^{(1)} = & \sqrt{p^{(1)}}\mathbf{F}^{(1)}\mathbf{R}_{\rm{cp}}^{(1)}\mathbf{H}^{(1)}\mathbf{x}_k^{(1)} +  
    \\ &\sqrt{p^{(2)}}\mathbf{F}^{(1)}\mathbf{R}_{\rm{cp}}^{(1)}\mathbf{H}^{(2)}\tildebf{x}_k^{(2)} + \boldsymbol{\omega}_k^{(1)},
    \label{eq:8}
    \end{split}
\end{equation}
where $$\boldsymbol{\omega}_k^{(1)} = \mathbf{F}^{(1)}\mathbf{R}_{\rm{cp}}^{(1)}\mathbf{w}_k.$$
Using (\ref{eq:2}) and (\ref{eq:5}) then gives
\begin{equation}
    \begin{split}
    \mathbf{y}_k^{(1)} = &\sqrt{p^{(1)}}{\mathbf{\Psi}}^{(1)}\mathbf{d}_k^{(1)} + \\ &\sqrt{p^{(2)}}\mathbf{F}^{(1)}\mathbf{R}_{\rm{cp}}^{(1)}\mathbf{H}^{(2)}\left[\mathbf{I}_2\otimes(\mathbf{A}_{\rm{cp}}^{(2)}({\mathbf{F}^{(2)})^{\rm{H}}})\right]\tildebf{d}_k^{(2)}\\
    &+ \boldsymbol{\omega}_k^{(1)}.
    \end{split}
    \label{eq:9}
\end{equation}
where the matrix ${\mathbf{\Psi}}^{(1)}$ is a square diagonal matrix with the DFT of the first column of $\mathbf{H}^{(1)}$, $[\psi^{(1)}_0, \psi^{(1)}_1, \dots, \psi^{(1)}_{N^{(1)}-1}]^{\rm{T}}$, on the main diagonal. The latter two terms in (\ref{eq:9}) are treated as noise and an estimate of symbol vector of user 1, $\mathbf{d}_k^{(1)}$, is recovered. This is then re-modulated and subtracted from \(\mathbf{r}_k\) and an estimate of the data symbol vector of user 2 is recovered.

\subsubsection*{Ordering 2 - SIC where user 2 is decoded first}

In this scenario the user with the larger subcarrier spacing is processed first. Since there are $q$ symbols from user 2 overlapping with a single symbol of user 1, these need to be isolated individually. To account for this, we introduce the matrix $$\mathbf{C}_{k,m}^{(2 \leftarrow 1)} = \left[\mathbf{0}_{L^{(2)}\times(m-1)L^{(2)}}, \mathbf{I}_{L^{(2)}}, \mathbf{0}_{L^{(2)}\times(q-m)L^{(2)}}\right],$$ which  isolates the $m$-th symbol of user 2 and the overlapping part of the symbol of user 1. The individual symbol from user 2 can then be recovered as
\begin{equation}
    \mathbf{y}_{k,m}^{(2)} = \mathbf{F}^{(2)}\mathbf{R}_{\rm{cp}}^{(2)}\mathbf{C}_{k,m}^{(2 \leftarrow 1)}\mathbf{r}_k,
    \label{eq:10}
\end{equation}
which, using (\ref{eq:3}), (\ref{eq:3.1})  and (\ref{eq:6}), yields 
\begin{equation}
    \begin{split}
    \mathbf{y}_{k,m}^{(2)} = &\sqrt{p^{(2)}}\mathbf{F}^{(2)}\mathbf{R}_{\rm{cp}}^{(2)}\mathbf{H}^{(2)}\mathbf{x}_{k,m}^{(2)} + \\ &\sqrt{p^{(1)}}\mathbf{F}^{(2)}\mathbf{R}_{\rm{cp}}^{(2)}\mathbf{C}_{k,m}^{(2 \leftarrow 1)}\mathbf{H}^{(1)}\mathbf{x}_k^{(1)} + \boldsymbol{\omega}_k^{(2)},
    \end{split}
    \label{eq:11}
\end{equation}
and using (\ref{eq:2}) then gives
\begin{equation}
    \begin{split}
    \mathbf{y}_{k,m}^{(2)} = &\sqrt{p^{(2)}}{\mathbf{\Psi}}^{(2)}\mathbf{d}_{k,m}^{(2)} +\\ &\sqrt{p^{(1)}}\mathbf{F}^{(2)}\mathbf{R}_{\rm{cp}}^{(2)}\mathbf{C}_k^{(2 \leftarrow 1)}\mathbf{H}^{(1)}\mathbf{A}_{\rm{cp}}^{(1)}({\mathbf{F}^{(1)})^{\rm{H}}}\mathbf{d}_k^{(1)} + \\
    &\boldsymbol{\omega}_k^{(2)}.
    \end{split}
    \label{eq:12}
\end{equation}
where ${\mathbf{\Psi}}^{(2)}$ is a diagonal matrix and the diagonal corresponds to the DFT of the first column of $\mathbf{H}^{(2)}$. 
Note that for Ordering 2, the $q$ overlapping symbols from user 2 need to be estimated, re-modulated, removed from \(\mathbf{r}_k\), and then the signal for user 1 can be recovered.

\section{Inter-Numerology Interference Analysis and Achievable Rate Derivation}
\subsection{Inter-Numerology Interference}
For Ordering 1 outlined previously, where user 1 is being decoded first, an INI matrix can be  calculated from the second term of (\ref{eq:9}) as

\begin{equation}
    \boldsymbol{\Gamma}^{(1 \leftarrow 2)} = \mathbf{F}^{(1)}\mathbf{R}_{\rm{cp}}^{(1)}\mathbf{H}^{(2)}\left[\mathbf{I}_q\otimes({\mathbf{A}}_{\rm{cp}}^{(2)}({{\mathbf{F}}^{(2)})^{\rm{H}}})\right],
    \label{eq:13}
\end{equation}
where \(\boldsymbol{\Gamma}^{(1 \leftarrow 2)}\) is an $N_{\rm{act}}^{(1)}\times{q}(N_{\rm{act}}^{(2)})$ matrix where the $(n,o_m)$-th element \({\Gamma}_{n,o_m}^{(1 \leftarrow 2)}\) contains the INI weight on subcarrier $n$ of user 1 from the the $o$-th subcarrier of the $m$-th overlapping symbol of user 2. For Ordering 2, where user 2 is decoded first, the INI matrix for the $m$-th symbol is calculated using the second term of (\ref{eq:12}) as
\begin{equation}
    \boldsymbol{\Gamma}_m^{(2 \leftarrow 1)} = \mathbf{F}^{(2)}\mathbf{R}_{\rm{cp}}^{(2)}\mathbf{C}_{k,m}^{(2 \leftarrow 1)}\mathbf{H}^{(1)}\mathbf{A}_{\rm{cp}}^{(1)}({\mathbf{F}^{(1)})^{\rm{H}}},
    \label{eq:14}
\end{equation}
where \(\boldsymbol{\Gamma}_m^{(2 \leftarrow 1)}\) is an $N_{\rm{act}}^{(2)}\times{N_{\rm{act}}^{(1)}}$ matrix where the $(n,o)$-th element \({\Gamma}_{n,o}^{(1 \leftarrow 2)}\) contains the INI weight on subcarrier $n$ of user 2 from the $o$-th subcarrier of user 1. For both orders, the INI matrix can be used to calculate the MSE interference on the victim user, 

\begin{equation}
    \boldsymbol{{\gamma}}^{(i \leftarrow j)} = \mathrm{diag}(\boldsymbol{\Gamma}^{(i \leftarrow j)}(\boldsymbol{\Gamma}^{(i \leftarrow j)})^{\mathrm{H}})
    \label{eq:15}
\end{equation}
which is a vector of length \(N_{\mathrm{act}}^{(i)}\) whose j-th element is equal to the MSE on the corresponding subcarrier of user $i$ due to interference from user $j$.

\subsection{Achievable Rates}
An approximation of the bandwidth used by user $i$ is given by
$B^{(i)} = N^{(i)}\Delta{f}^{(i)}.$
The cyclic prefix converts the channel into $N^{(i)}$ parallel subchannels. Assuming user $i$ is decoded first in the SIC scheme and that the CP is correctly discarded at the receiver, the instantaneous achievable rate for user $i$ is given by

\begin{equation}
    R^{(i)} = \sum_{n=1}^{N^{(i)}}\log_2\left(1+\mathrm{SINR}^{(i)}_n\right) \rm{bits/symbol},
    \label{eq:16}
\end{equation}
which can also be written as
\begin{equation}
    R^{(i)} = \frac{B}{N^{(i)}}\sum_{n=1}^{N^{(i)}}\log_2\left(1+\mathrm{SINR}^{(i)}_n\right) \rm{bits/second},
    \label{eq:18}
\end{equation}
where \(\mathrm{SINR}^{(i)}_n\) denotes the instantaneous signal-to-interference-plus-noise ratio (SINR) of the \(n\)-th subcarrier, which is given by
\begin{equation}
    \mathrm{SINR}^{(i)}_n = \frac{p^{(i)}|\psi^{(i)}_n|^2}{{p^{(j)}\boldsymbol{\gamma}}^{(i\leftarrow j)}_n+N_0}.
    \label{eq:19}
\end{equation}
User $j \ne i$ is decoded second and does not experience any interference from user $i$; therefore, its achievable rate is simply given by
\begin{equation}
    R^{(j)} = \frac{B}{N^{(j)}}\sum_{n=1}^{N^{(j)}}\log_2\left(1+\mathrm{SNR}^{(j)}_n\right) \rm{bits/second},
    \label{eq:20}
\end{equation}
where $$\mathrm{SNR}^{(j)}_n = \frac{p^{(j)}|\psi^{(j)}_n|^2}{N_0}$$ is the signal-to-noise ratio of the \(n\)-th subcarrier of user $j$.
The achievable rates for each user are given by
   \begin{equation}
   R^{(i)} = \frac{B}{N^{(i)}}\sum_{n=1}^{N^{(i)}}\log_2\left(1+\frac{p^{(i)}|\psi^{(i)}_n|^2}{p^{(j)}\gamma^{(i\leftarrow j)}_n+N_0}\right),
   \label{eq:21}
   \end{equation}
and
    \begin{equation}
    R^{(j)} = \frac{B}{N^{(j)}}\sum_{n=1}^{N^{(j)}}\log_2\left(1+\frac{p^{(j)}|\psi^{(j)}_n|^2}{N_0}\right).
    \label{eq:22}
    \end{equation}

\section{Results and Discussion}
\begin{figure}[b]
    \centering
    \includegraphics[width=\columnwidth]{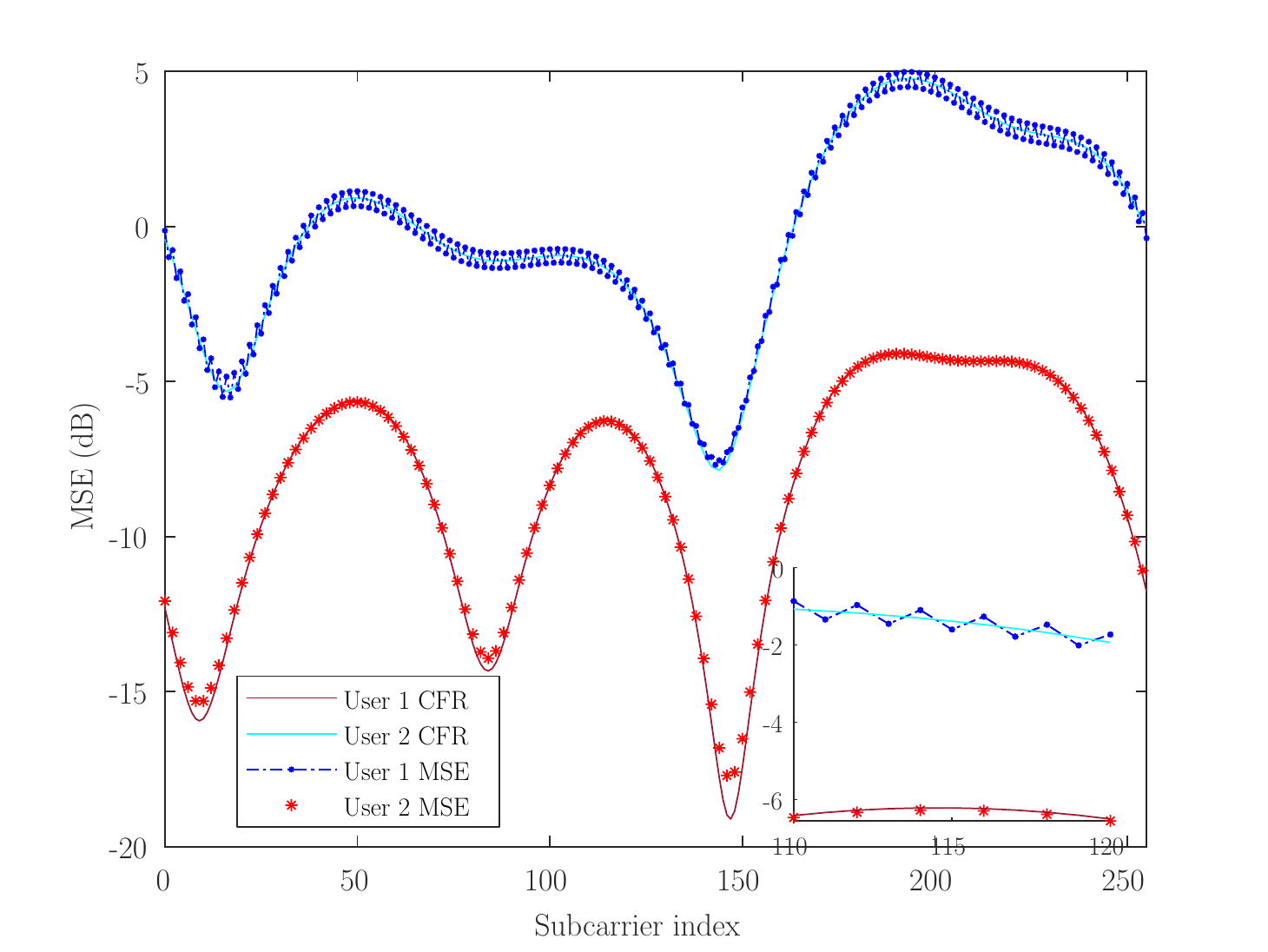}
    \caption{\label{fig:2} MSE and CFR for user 1 and 2.}
\end{figure}
This section shows the analytical results obtained from the system model and analysis outlined above. For all the results shown in this section, all available subcarriers for both users are active, i.e. $ N_{\rm{act}}^{(i)} = N^{(i)}$ for each user $i \in \{ 1,2 \}.$ Table \ref{tab:table 1} lists the numerologies that we have used with their number of subcarriers and  CP lengths, which are in line with 5G NR specifications \cite{3gpp_TS38211}.

\begin{table}[h]
    \caption{5G NR Numerologies}
    \centering
    \begin{tabular}{|c|c|c|}
        \hline
        \hline
         Numerology & No. of Subcarriers & CP Length \\
         \hline
         \hline
         0 & 4096 & 288 \\
         \hline
         1 & 2048 & 144 \\
         \hline
         2 & 1024 & 72 \\
         \hline
         3 & 512 & 36 \\
         \hline
         4 & 256 & 18 \\
         \hline
         5 & 128 & 9 \\
         \hline
    \end{tabular}
    \label{tab:table 1}
\end{table}

\subsection{Inter-Numerology Interference}

\begin{figure}[b!]
    \begin{subfloat}
    \centering
    \includegraphics[width=\columnwidth]{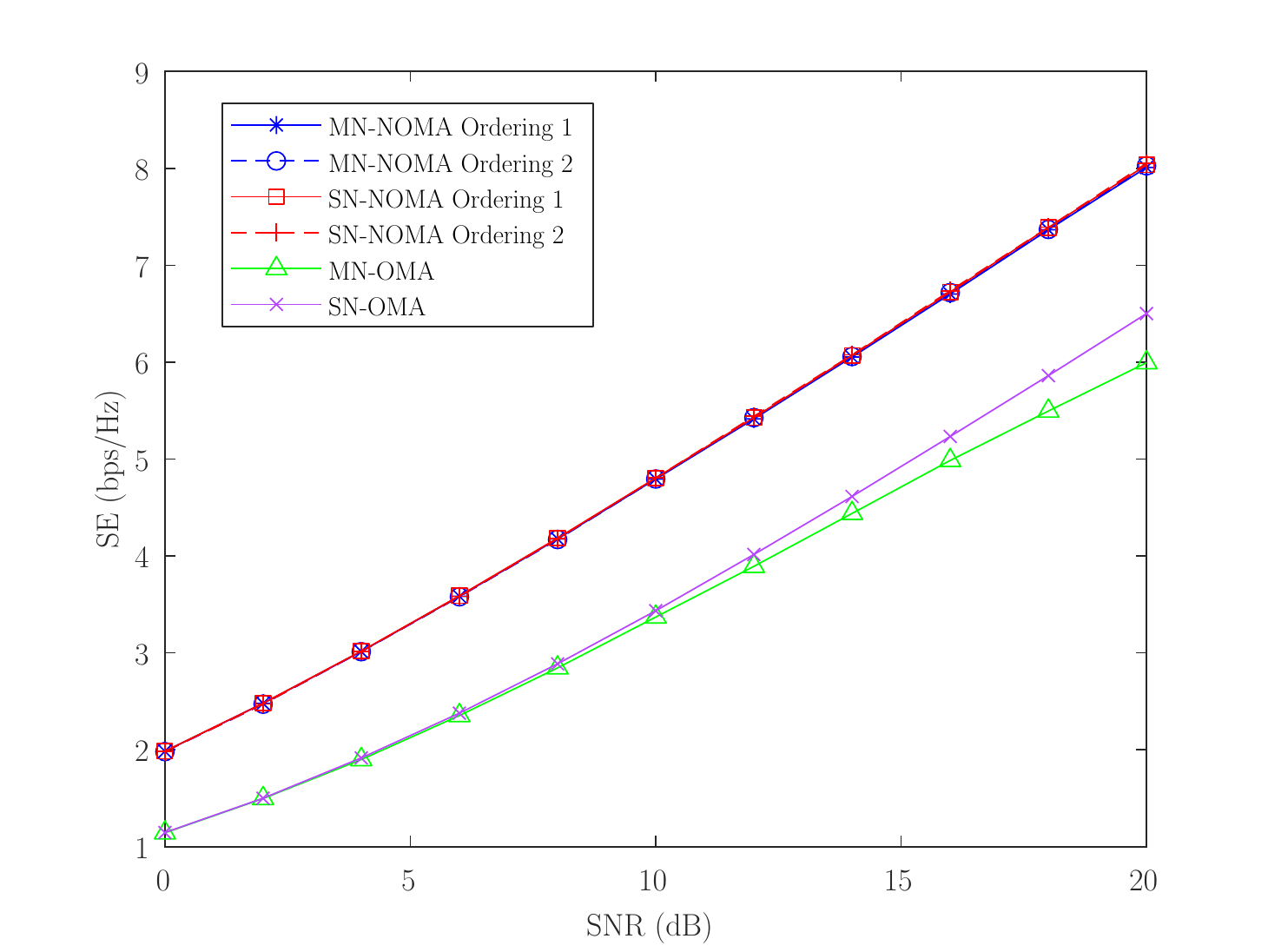}
    \caption{\label{fig:4} Spectral efficiency for MN-NOMA, SN-NOMA and MN-OMA for different levels of SNR. Both users use the EPA channel model.}
    \end{subfloat}
    \begin{subfloat}
    \centering
    \includegraphics[width=\columnwidth]{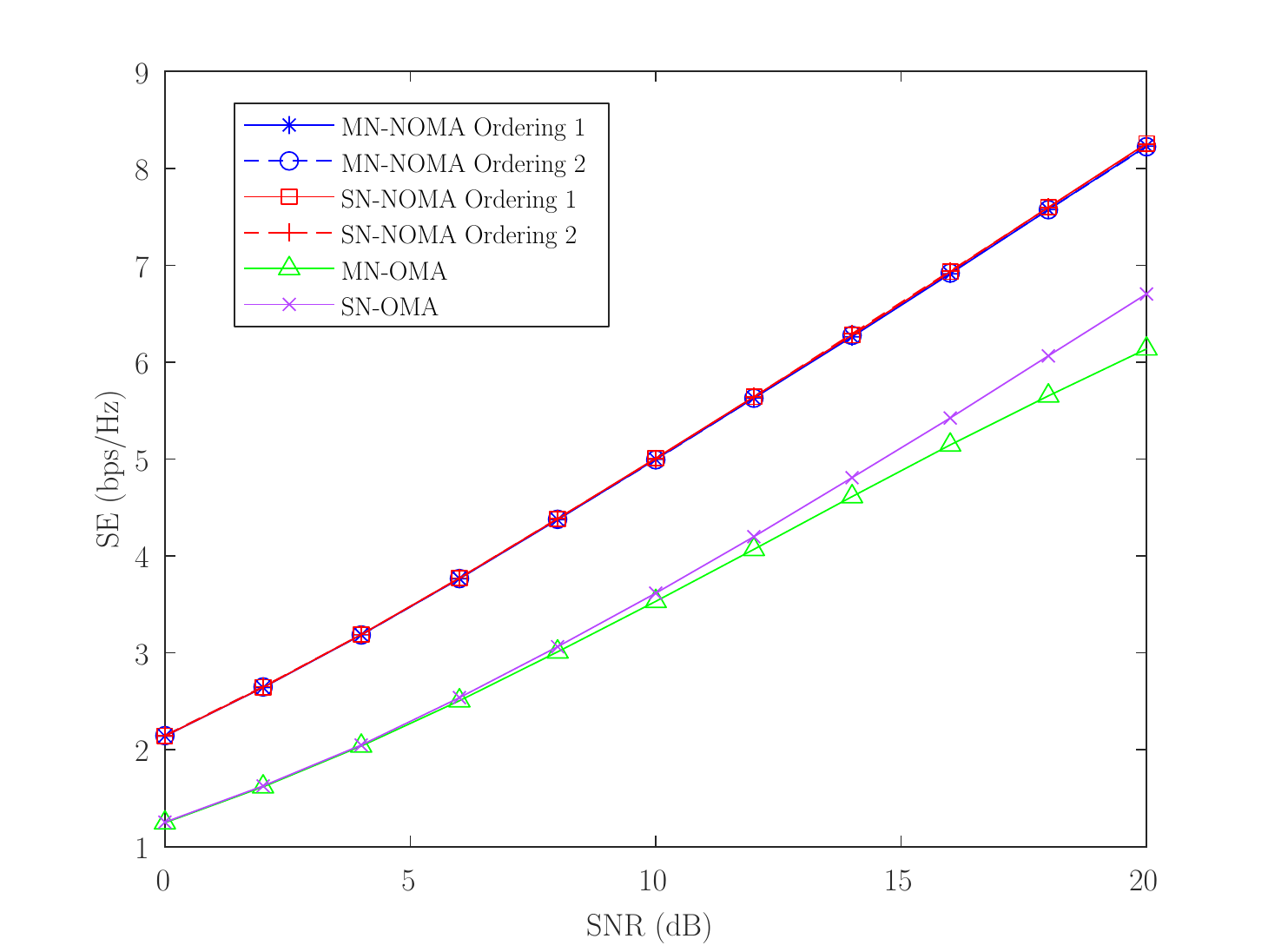}
    \caption{\label{fig:5} Spectral efficiency for MN-NOMA, SN-NOMA and MN-OMA for different levels of overall SNR. Users 1 and 2 use the EPA and EVA channel model, respectively.}
    \end{subfloat}
\end{figure}

 \begin{figure}[b!]
    \begin{subfloat}
    \centering
    \includegraphics[width=\columnwidth]{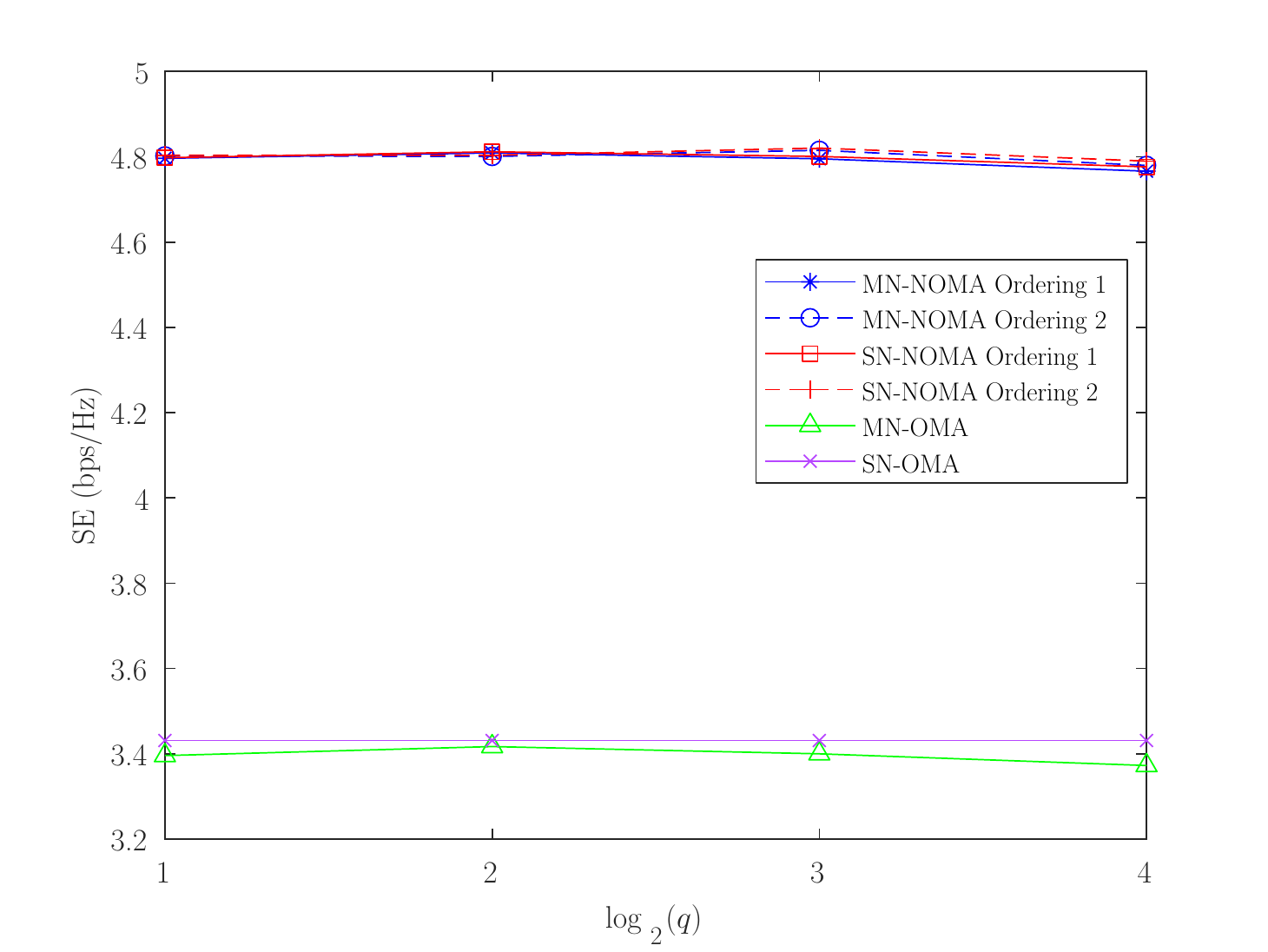}
    \caption{\label{fig:6} Spectral efficiency for MN-NOMA and SN-NOMA for different values of $q = N^{(1)}/N^{(2)}$. Both users use the EPA channel model.}
    \end{subfloat}
    \begin{subfloat}
    \centering
    \includegraphics[width=\columnwidth]{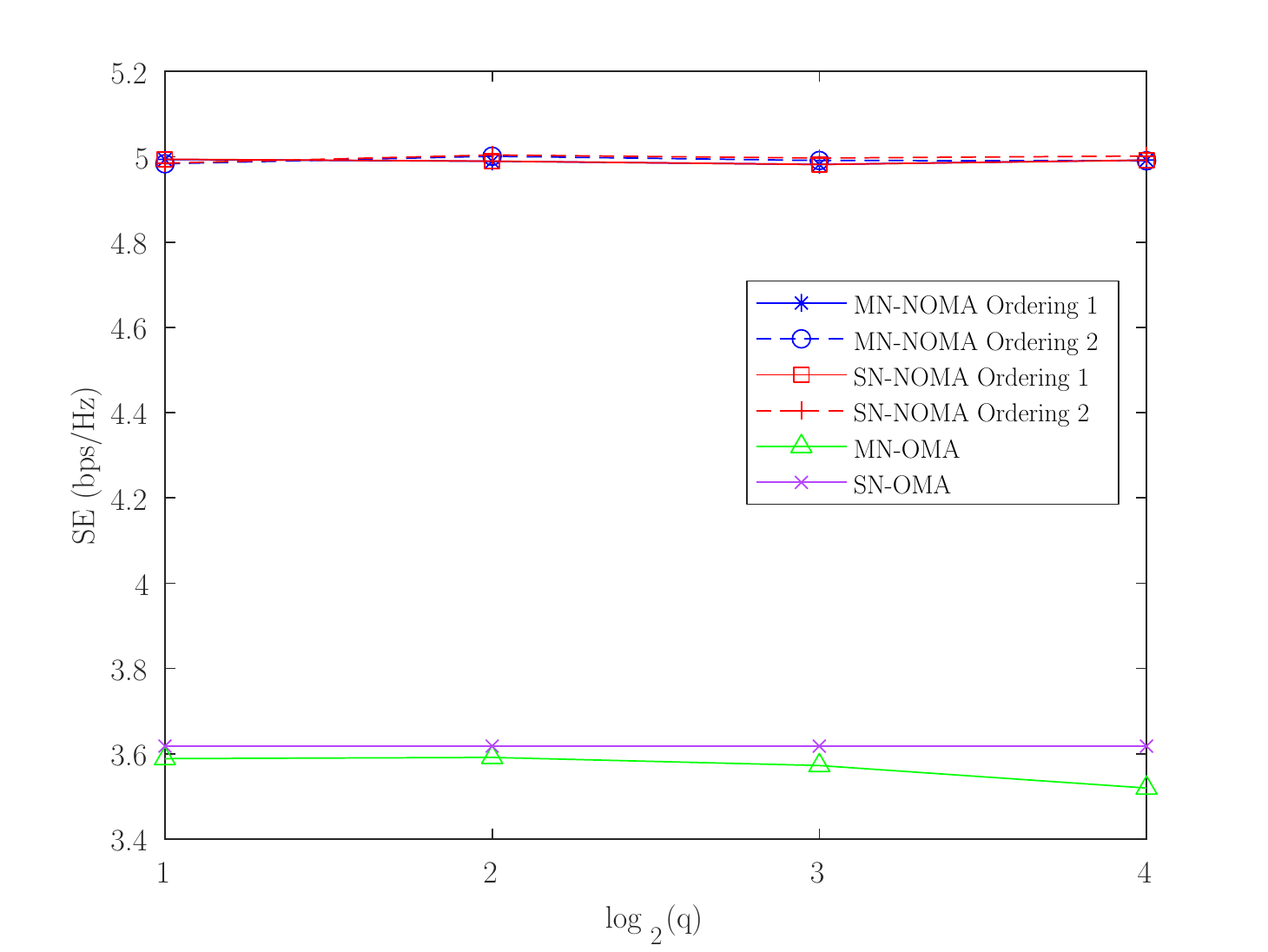}
    \caption{\label{fig:7} Spectral efficiency for MN-NOMA and SN-NOMA for different values of $q = N^{(1)}/N^{(2)}$. Users 1 and 2 use the EPA and EVA channel model, respectively.}
    \end{subfloat}
\end{figure}

For the results in this subsection, user 1 used numerology 4 and user 2 used numerology 5 as outlined in Table \ref{tab:table 1}. Each user transmits over a particular fading channel generated using the Extended Pedestrian A (EPA) channel model \cite{3gpp_TS136116}.

Fig. \ref{fig:2} shows the channel frequency response (CFR) and the MSE of each user. It can be seen that for each user, the MSE follows the shape of the CFR curve of the other user. However, the MSE of user 1 generally oscillates around the CFR of user 2 from subcarrier to subcarrier. This is due to the orthogonality between a user 1 subcarrier and the adjacent subcarriers of user 2 being broken, causing even subcarriers to experience more interference and the odd subcarriers to experience less interference. The MSE for user 2 does not experience this phenomenon. This is to be expected as the orthogonality between a subcarrier of user 2 and the adjacent subcarriers of user 1 is not broken. This shows there is an inherent asymmetry between the nature of interference each user experiences.

\subsection{Spectral Efficiency}
In this subsection we evaluate the spectral efficiency of MN-NOMA and compare this to SN-NOMA, multinumerology OMA (MN-OMA) and single numerology OMA (SN-OMA). The evaluation of the achievable rate of each user is performed via Monte Carlo simulation over a large number of random channel instances. For each channel instance in the Monte Carlo simulation, the power allocation for MN-NOMA and for SN-NOMA is optimized using an exhaustive search algorithm to maximize the sum rate for Ordering 1 and Ordering 2. The power allocation for both users in MN-OMA and SN-OMA is set as $p^{(1)} = p^{(2)} = P/2.$

In Fig. \ref{fig:4} and Fig. \ref{fig:5} we compare the average spectral efficiency for MN-NOMA, SN-NOMA and MN-OMA for different levels of SNR. For these simulations, user 1 uses numerology 4 and user 2 uses numerology 5 from Table \ref{tab:table 1}, i.e. $q = 2$. In Fig \ref{fig:4} both users use the EPA channel model while in Fig. \ref{fig:5} user 1 uses the EPA channel model and user 2 uses the extended vehicular A (EVA). It can be seen from these figures that for both Orderings 1 and 2 the average spectral efficiency for MN-NOMA with optimized power allocation is extremely close to that of SN-NOMA. It can also be seen that for both Orderings, the spectral efficiency of MN-NOMA is superior to that of an MN-OMA system using the same numerologies. This shows that MN-NOMA offers the same spectral efficiency benefits as SN-NOMA over a range of different SNRs for this pair of numerologies.

In Fig. \ref{fig:6} and Fig. \ref{fig:7} we compare the average spectral efficiency for MN-NOMA, SN-NOMA and MN-OMA for different values of the ratio $q$. For these simulations, user 1 uses numerology 1 and the numerology of user 2 is varied from numerology 2 to numerology 5. In Fig. \ref{fig:6} both users used EPA and in Fig. \ref{fig:7} user 2 is changed to EVA. A constant $\mathrm{SNR} = 10\mathrm{dB}$ is used. It can be seen from these figures that, for different values of $q$, the average spectral efficiency for MN-NOMA is superior to that of MN-OMA and is still very close to that of SN-NOMA.

These results show that while the nature of the interference experienced by users in MN-NOMA may be different to that of SN-NOMA, we can still achieve the flexibility benefits of multinumerology systems while also gaining the spectral efficiency benefits of NOMA. The results also show that despite this asymmetry in how each user experiences interference from the other in MN-NOMA, on average the spectral efficiency is the same for Ordering 1 and Ordering 2 under optimized power allocation. 

\section{Conclusion}
This paper develops a generalized system model for multinumerology NOMA with users transmitting over multipath fading channels. This model is used to derive analytical expressions for the interference induced on one user by another, depending on the order of SIC. These analytical expressions are used to derive expressions for the instantaneous achievable rates of each user. Numerical results show that the interference experienced by a user with narrower subcarrier spacing is different in MN-NOMA than in SN-NOMA. The spectral efficiency results provided then show that despite this change in interference experienced by the users, MN-NOMA has very similar achievable rates to SN-NOMA. This shows that the MN-NOMA system can gains the flexibility of being able to accommodate users with different numerologies while also gaining the spectral efficiency benefits associated with NOMA.
\newpage
The work presented in this paper provides plenty of potential future research. The power allocations used throughout this paper are constant across all subcarriers and there is scope for further optimization via using the water filling algorithm to take advantage of the fact that the MSE for the user with narrower subcarrier spacing oscillates from one subcarrier to another. Furthermore, the model presented her assuming perfect synchronicity in time and frequency; it would be interesting to investigate whether the demonstrated gains and system flexibility can be maintained also in asynchronous cases of multinumerology NOMA.

\section*{Acknowledgments}
This publication has emanated from research supported in part by a research grant from Science Foundation Ireland (SFI) and is co-funded under the European Regional Development Fund under Grant Number 13/RC/2077.
This work was also supported in part by the U.K. Engineering and Physical Sciences Research Council (EP/S02476X/1).


\bibliographystyle{IEEEtran}
\bibliography{mixed_num_NOMA_bib}
\end{document}